\documentclass[superscriptaddress,twocolumn,aps,
showpacs]{revtex4-1}
\usepackage{graphicx}
\usepackage{amsbsy,amssymb,amsmath,bm,ulem}

\usepackage{graphicx}
\usepackage{dcolumn}
\usepackage{bm}
\usepackage{hyperref}

\usepackage{color}
\usepackage{csquotes} 

\begin{document}

\preprint{APS/123-QED}

 
 \title{ 
Active control of thermomagnetic avalanches in superconducting Nb films with tunable anisotropy
}

\author{D. Carmo}
\affiliation{Departamento de F\'{i}sica, Universidade Federal de S\~{a}o Carlos, 
13565-905, S\~{a}o Carlos, SP, Brazil}

\author{F. Colauto}
\affiliation{Departamento de F\'{i}sica, Universidade Federal de S\~{a}o Carlos, 
13565-905, S\~{a}o Carlos, SP, Brazil}

\author{A. M. H. de Andrade}
\affiliation{Instituto de F\'{i}sica, Universidade Federal do Rio Grande do Sul, 
91501-970, Porto Alegre, RS, Brazil}

\author{A. A. M. Oliveira}
\affiliation{Instituto Federal de Educa\c{c}\~{a}o, Ci\^{e}ncia e Tecnologia 
de S\~{a}o Paulo, 13565-820, S\~{a}o Carlos, SP, Brazil}

\author{W. A. Ortiz}
\affiliation{Departamento de F\'{i}sica, Universidade Federal de S\~{a}o 
Carlos, 13565-905, S\~{a}o Carlos, SP, Brazil}

\author{Y. M. Galperin}
\affiliation{Department of Physics, University of Oslo, P. O. Box 1048 
Blindern, 0316 Oslo, Norway}

\author{T. H. Johansen}
\affiliation{Department of Physics, University of Oslo, P. O. Box 1048 
Blindern, 0316 Oslo, Norway}
\affiliation{Institute for Superconducting and Electronic Materials, 
University of Wollongong, Northfields Avenue, Wollongong, NSW 2522, Australia}

\date{\today}

\begin{abstract}
Active triggering and manipulation of ultrafast flux dynamics in superconductors are demonstrated in films of Nb. Controlled amounts of magnetic flux were injected from a point along the edge of a square sample, which at 2.5 K responds by nucleation of a thermomagnetic avalanche.
Magneto-optical imaging was used to show that when such films are cooled in the presence of in-plane magnetic fields they become anisotropic, and the morphology of the avalanches change systematically, both with the direction and magnitude of the field.
The images reveal that the avalanching dendrites consistently bend towards the direction perpendicular to that of the in-plane field.
The effect increases with the field magnitude, and at 1.5 kOe the triggered avalanche becomes quenched at the nucleation stage.
The experimental results are explained based on a theoretical model for thermomagnetic avalanche nucleation in superconducting films, and by assuming that the frozen-in flux generates in-plane anisotropy in the film thermal conductance.
The results demonstrate that applying in-plane magnetic fields to film superconductors can be a versatile external tool for controlling their ultrafast flux dynamics.
\end{abstract}

\maketitle
\section{Introduction}

In type-II superconducting films anisotropy often causes a preferred direction for motion of the vortices present in the material\cite{berghuis_intrinsic_1997, dobrovolskiy17}.
The anisotropic behavior can have many different origins, e.g., the presence of material microstructures such as twin and anti-phase boundaries\cite{turchinskaya_direct_1993, jooss_pinning_2000}, planar microdefects \cite{cuche_influence_1996}, misoriented substrates\cite{polyanskii_magneto-optical_2005, qviller_quasi-one-dimensional_2012}, asymmetric pinning potentials\cite{he_magneto-optical_2009},  columnar defects\cite{schuster_critical-current_1995, leonhardt_influence_2000}, and patterned arrays of holes\cite{tsuchiya_anisotropic_2014}.
In all these cases the anisotropy is  due to permanent characteristics of the sample.
A quite different way to produce anisotropic flux dynamics is by applying crossing magnetic fields \cite{indenbom_anisotropy_1994}.
This extrinsic approach has the important features of being fully reversible and easily tunable \cite{schuster_discontinuity_1995, vlasko-vlasov_crossing_2016}.
Quite recently a similar approach was demonstrated  in a hybrid system where an array of stripes of soft magnetic material was placed on top a superconducting film \cite{vlasko-vlasov_manipulating_2017}.

Anisotropic superconductors are today considered promising for development of active fluxonics devices such as diodes \cite{de_souza_silva_dipole-induced_2007} and triodes \cite{vlasko-vlasov_triode_2016} based on controlling the vortex matter dynamics\cite{crete_devices_2002}.
However, the intrinsic anisotropy can be insufficient to generate the required preferential vortex motion. In that case, adding an external in-plane magnetic field can serve to enhance the maximum anisotropy of the superconducting film, and do this in a tunable manner.
Conversely, if isotropic properties are desired in an anisotropic superconducting film, an in-plane field could be applied to compensate for the intrinsic anisotropy.

In this work magneto-optical imaging (MOI) was used to investigate flux penetration in Nb films, which in zero in-plane field behave fully isotropic. 
When freezing in  in-plane magnetic fields  the anisotropy grows dramatically, and is found to  follow a cubic dependence on the frozen-in field. 
A local flux injector was used to explore how one can by varying the  anisotropy dramatically change the characteristics of thermomagnetic avalanches, which we 
trigger by passing a current pulse in the injector.
The experimental results are discussed, and the main characteristics are explained based on a theoretical model for the thermomagnetic instability of  superconducting films coupled thermally to their substrate.

\begin{figure}[b]
  \centering
  \includegraphics[width=8cm]{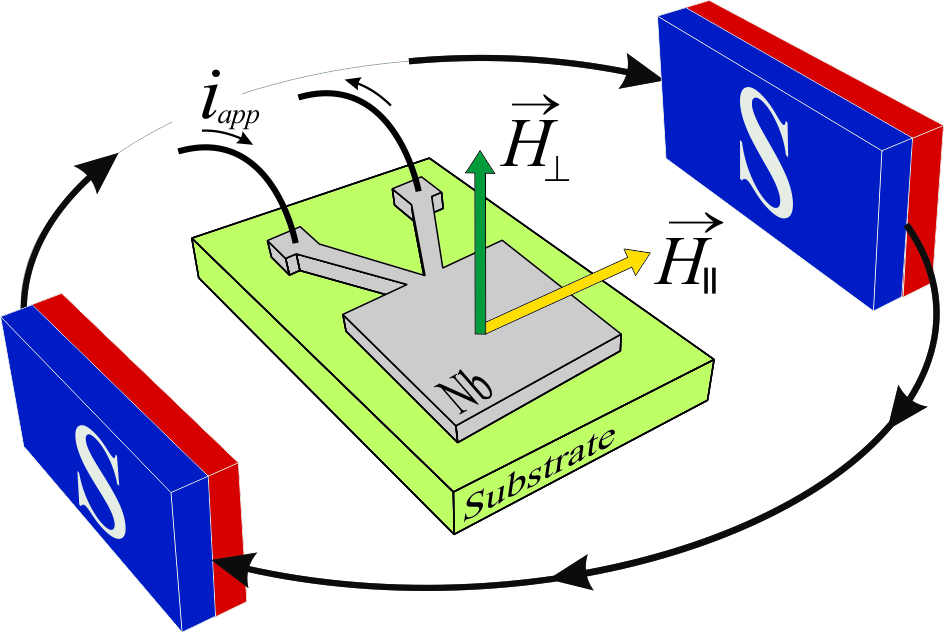}
   \label{fig:drawing}
  \caption{Configuration of the experimental setup where the sample is exposed to tunable external magnetic fields, $H_{\perp}$ and $H_{\parallel}$, 
and an applied current  $i_{\rm \textit{app}}$.}
\end{figure}

\section{Materials and Methods}
Samples were prepared from 200~nm thick Nb films grown on Si (100) substrates by magnetron sputtering in a UHV system with base pressure less than $2 \times 10^{-8}$ Torr. 
The films have a critical temperature of $T_c$ = 9.1~K, and patterning was performed using optical lithography.
Two different samples were made, where one was shaped as a plain square with sides $a =2.5$~mm, serving as reference sample.
The other is an identical square film extended by a V-shaped structure, see Fig.~1, where the top of the ``V" constitutes a pair of contact pads allowing to apply electrical current pulses, $i_{\rm app}$. 
The current will then pass through the large square only via a small region where the two legs of the ``V" merge, resulting in a sharp turn of the current flow.
The associated magnetic field becomes highly focused between these meeting legs, and allows to make local injections of  magnetic flux into the square area.\cite{carmo_controllable_2016}

The setup allows also for external magnetic fields to be applied.
This is accomplished by placing a resistive coil around the cryostat (not shown in Fig.~1), thus providing a perpendicular field $H_{\perp}$ applied to the sample.
In addition, an in-plane field $H_{\parallel}$ is generated by a pair of permanent magnets mounted on a specially designed stage, allows to apply various fields $H_{\parallel}$.
The field magnitude is controlled by  the magnet-magnet separation, and the orientation  is set by rotating the whole stage.

To observe how the magnetic flux penetrates into the sample we use a magneto-optical imaging (MOI) setup where a thin plate of Faraday rotating Bi-substituted ferrite-garnet film with in-plane magnetization serves as sensor \cite{helseth_faraday_2001}.
The sensor plate is placed directly on top of the superconducting film. 
When viewed in a polarized light microscope with crossed polarizers, the  image brightness represents a direct map of the perpendicular flux density 
distribution,\cite{johansen_direct_1996, vlasko-vlasov_magneto-optical_1999, jooss_magneto-optical_2002} both inside and outside the sample area.
The objective lens used in this work was a 5X MPlanFL Olympus, and images were recorded using a Qimaging Retiga 4000R CCD camera. 
Its pixel scale was calibrated based on a Pelcotec TM CDMS Standard.
The image analysis was carried out using the Image J 1.48 software.
The current pulses were generated by a Keithley-2635 current source.

\section{Results and Discussion}
\subsection{Flux penetration at 7~K}

To determine the field-induced anisotropy characteristics of our Nb films separate experiments were carried out on a plain square sample.
Shown in Fig.~2 are magneto-optical images of the sample recorded after initial cooling to 7~K.
At this temperature flux penetrates quite smoothly, i.e., without any observable intermittency as the applied perpendicular field is increased. 
Panel (a) shows the flux density distribution when the field reached $H_{\perp} = 50$~Oe, which caused flux to penetrate the entire sample area.
The perimeter of the sample is here seen as the bright contour due to the diamagnetic property of the superconductor. 
The dark diagonal lines show where the planar shielding current makes sharp turns to adapt to the critical-state conditions in a square geometry.\cite{schuster_observation_1994}
In panel (a), where no in-plane field was applied, the behavior is fully isotropic, as evidenced by the equal size of the 4 triangular domains where the critical current flows with the same density, $J_c$, but in different directions.

Panel~(b) of Fig.~2 shows the result of the same experimental procedure creating the image in (a), except here an in-plane field of magnitude $H_{\parallel}$ = 0.7~kOe was applied during the initial cooling of the sample.
The direction of $H_{\parallel}$ is indicated by the arrow in the figure, and this field  was maintained also when the perpendicular field $H_{\perp}$ = 50~Oe  was reapplied.
One sees here that the flux penetration pattern changed substantially, as a new horizontal dark line appears in the central part of the image.
Both domains where the current flow is transverse to $H_{\parallel}$ have now grown in size, while  the domains where the current flow is aligned (parallel  or anti-parallel) with $H_{\parallel}$ have been reduced.
Evidently, the in-plane field generates a significant anisotropy in the critical current flow.\cite{vlasko-vlasov_crossing_2016}

\begin{figure}[t]
  \centering
  \includegraphics[width=8.0cm]{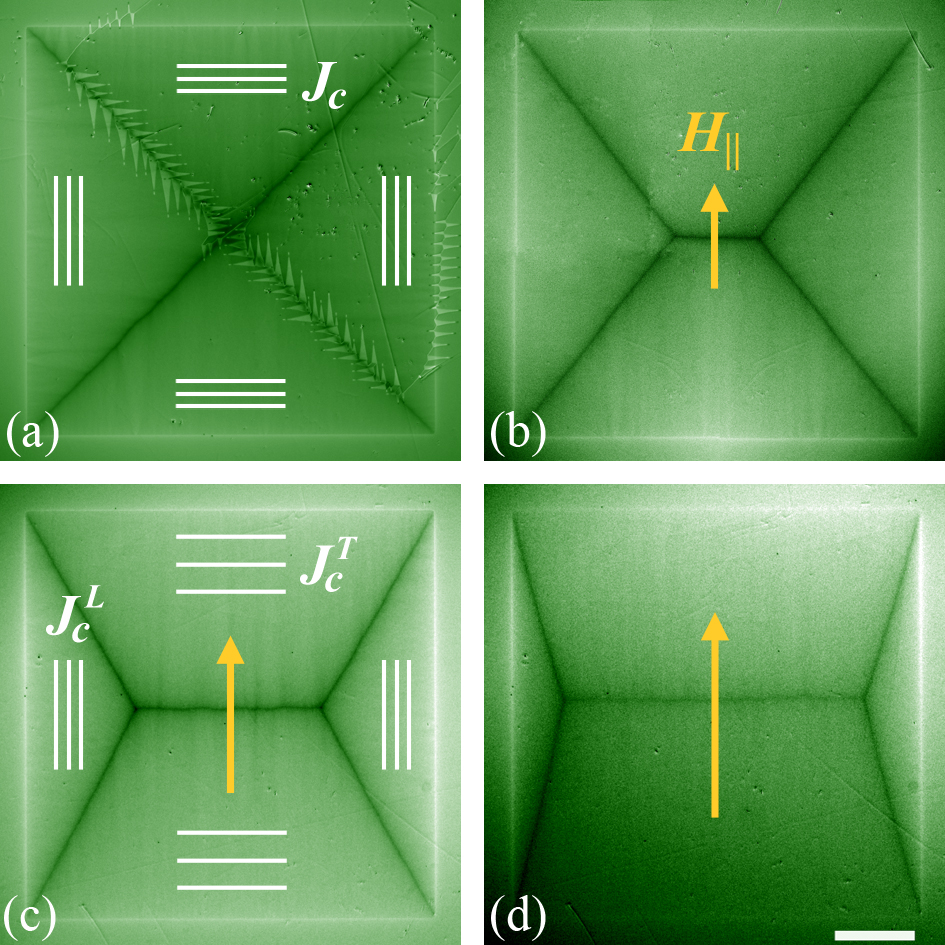}
  \label{fig:square7K}
  \caption{Magneto-optical images of a square Nb film at $T$~=~7~K, when applying a perpendicular magnetic field $H_{\perp}$~=~50~Oe resulted in full flux penetration. In panel (a) and (c) the 4 sets of line segments indicate stream lines of the shielding current flow in each domain. 
In panels (b)-(d) the sample was initially cooled in the presence of in-plane fields (see arrows) of  $H_{\parallel}$ = 0.7,  1.0, and 1.5 kOe, respectively. The scale bar at the bottom right is 500~$\mu$m}
 \end{figure}

Presented in Fig.~2 (c) and (d) are the results of similar experiments carried out with even larger in-plane fields, $H_{\parallel}~=~1.0$ and 1.5~kOe, respectively.
The anisotropy clearly continues to increase with the magnitude of the in-plane applied field.
  
\begin{figure}[t]
  \centering
  \includegraphics[width=8cm]{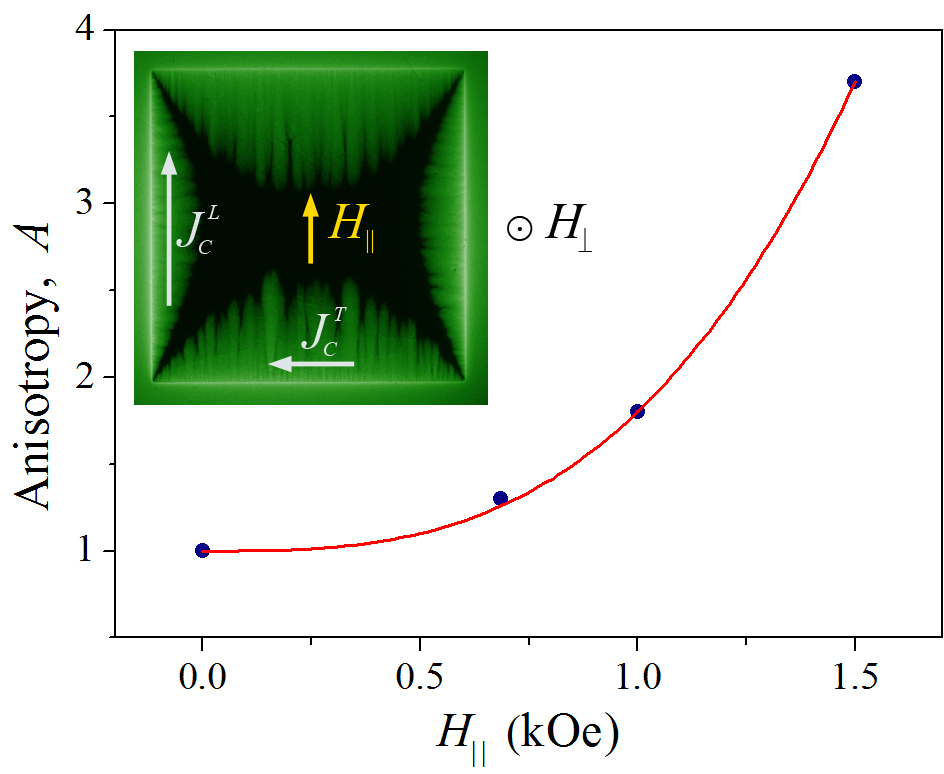}
  \label{fig:power}
  \caption{Field-induced anisotropy,  $A = J_c^{L}/J_c^{T}$, measured at 7~K as function of the initially frozen-in in-plane field, $H_{\parallel}$.  
  The full curve shows a fitted cubic power-law. Insert: Magneto-optical image of the in-plane field-cooled Nb film ($H_{\parallel}~=$ 1~kOe) exposed to a $H_{\perp} =$ 20~Oe perpendicular field.}
\end{figure}

To quantify this easily tunable anisotropy we adopt the critical-state model, which predicts  exactly the type of flux penetration patterns seen in Fig.~2
provided the magnitude of the critical current density is different in the two directions, {\it (i)} along, and {\it  (ii)} transverse to the in-plane field.
Letting $J_c^{L}$ and  $J_c^{T}$ denote these respective critical sheet currents, the anisotropy can be extracted directly from the images in Fig.~2 using that,
\begin{equation}
A \equiv J_c^{L}/J_c^{T} = (1 - x/a)^{-1} \, ,
\end{equation}
where $x$ is the length of the dark horizontal line.
From the panels (a)-(d)  one then finds the following values for the anisotropy, $ A = 1.0 , 1.3, 1.8, 3.7$.
These data  are plotted in Fig.~3 as function of the in-plane field. Included in the plot is  also a curve representing  the form,
\begin{equation}
 A   = 1+ c~ H_{\parallel}^3  \, ,
\end{equation}
with $c = 8 \cdot 10^{-10}$~Oe$^{-3}$, which gives an excellent fit to the experimental data. 

The anisotropy is evident in magneto-optical images also before full flux penetration is reached. 
An example is shown in the insert of Fig.~3.
Here the film was initially cooled in $H_{\parallel}$ =1~kOe, and then a perpendicular field of $H_{\perp}$ = 20~Oe was applied.
In addition to the clear difference in the flux penetration depth from the two pairs of opposite edges, the image also reveals that the penetration has a clear filamentary character in the direction parallel to the frozen-in flux.
This suggests that the motion of incoming perpendicular vortices is facilitated and  guided by the in-plane vortices already present in the film, an observation fully consistent with MOI results reported recently \cite{colauto_anisotropic_2017}.
In that work it was  found that in terms of critical current densities the in-plane magnetic field has little influence on the magnitude of $J_c^{L}$. 
Thus, the field-induced anisotropy is essentially due to $J_c^{T}$ becoming smaller with increasing $H_{\parallel}$, as indicated by the spacing between the current stream lines in Fig.~2.
 
\subsection{Local flux injections at 2.5~K}

Consider now the outcome of activating the flux injector at $T= $~2.5~K, a temperature where Nb films are thermomagnetically unstable and avalanches of magnetic flux  easily occur \cite{duran95, altshuler_Nb_2004}.
Figure~4  presents magneto-optical images of the Nb film where the injector is seen to be located at the lower sample edge. 
After the initial cooling, a small perpendicular field, $H_{\perp}$ = 5~Oe, was  turned on only to make the film edges visible for MOI (see bright contour).
Then, a current pulse of magnitude $i_{app}~=~1$~A and duration 200~ms was applied to the injector.

Panel (a) of Fig.~4 shows the resulting flux distribution when no in-plane magnetic field was frozen-in prior to the current pulse.
From the image one sees that the pulse caused here a substantial amount of  flux to invade the film in the form of a dendritic avalanche with four branches of similar lengths.
The branches are here all rooted at the point where the two legs of the injector meet, which is where the local magnetic field is largest during the current pulse.
Thus, the branches emanate from the nucleation site expected in the present geometry.

In panel (b) the same current pulse was applied after the sample this time was initially cooled in the presence of an in-plane field, $H_{\parallel}$ = 1.5~kOe  directed as shown by the arrow.
Evidently, this field-cooling prevented a similar avalanche to occur.

\begin{figure}[t]
  \centering
  \includegraphics[width=8cm]{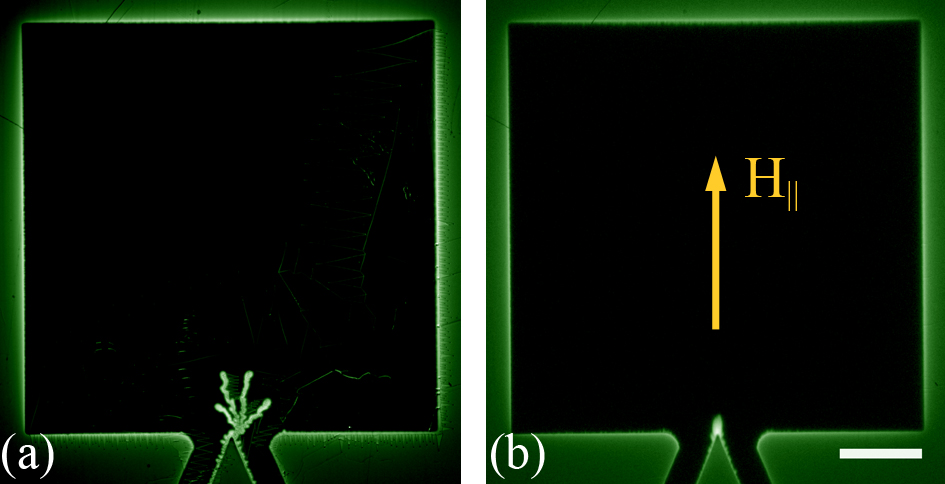}
  \label{fig:ava-quench}
  \caption{Magneto-optical images of the Nb film with a flux injector at the lower edge.
  In (a) the film was initially zero-field-cooled to 2.5~K before a current pulse triggered a dendritic flux avalanche.
 In (b) an in-plane field,  $H_{\parallel}$ = 1.5~kOe  (see arrow) was applied during the cooling before applying the same current pulse.
 The scale bar is 0.5~mm long. 
  }
\end{figure}

\subsection{Quenching of dendrites}

To explain this striking quenching effect caused by the in-plane magnetic field we make use of theoretical results obtained for  the thermomagnetic instability in superconducting films coupled thermally to their substrate.\cite{denisov06-1}
Those analyses consider a film strip of width $2w$ and thickness $d\ll w$ placed in an increasing perpendicular magnetic field  creating a critical state in the flux-penetrated region near the edges.
By solving the Maxwell and  thermal diffusion equations with proper boundary conditions, it was shown that for small fields there is no solution for perturbations growing with time, implying a stable situation.
As the perpendicular field increases the flux distribution can become unstable, and result in abrupt penetration of magnetic flux in the form of a dendritic structure.

\begin{figure}[t]
  \centering
  \includegraphics[width=8cm]{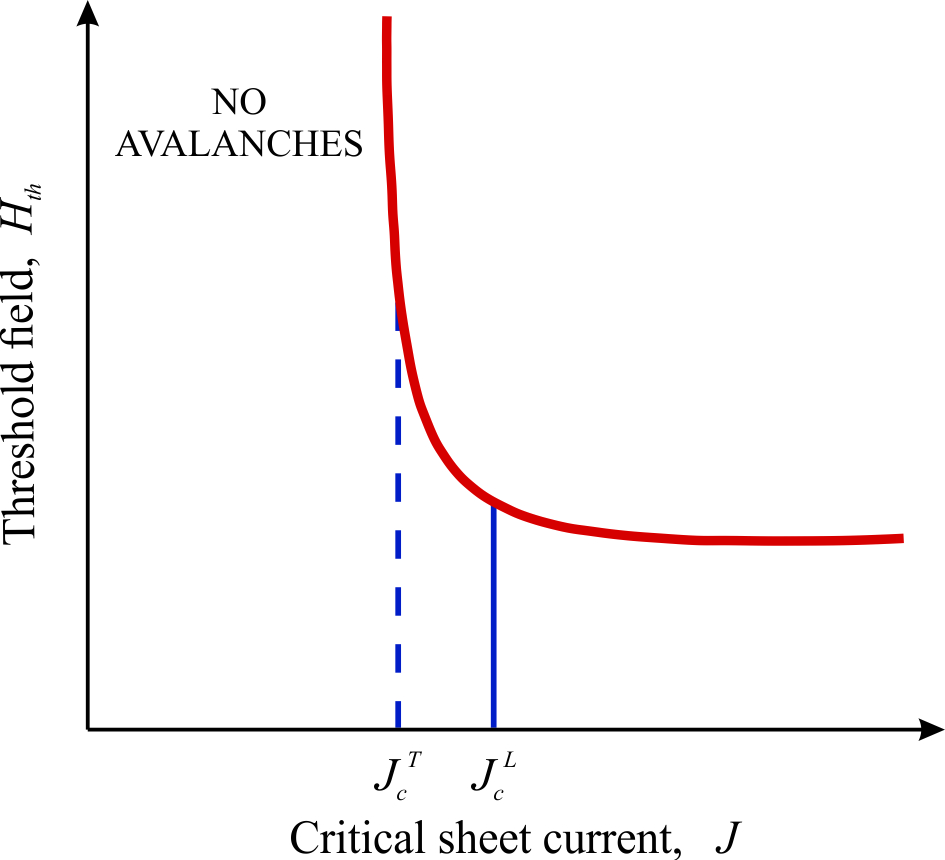}
  \label{Hth-j}
  \caption{Generic curve for threshold applied perpendicular field for the onset of  thermomagnetic  avalanche activity in superconducting films versus their critical sheet current.}
\end{figure}

Within this model, a  superconducting film of thickness, $d$, becomes thermomagnetically unstable when the flux penetration depth, $\ell$, reaches the value given by,\cite{denisov06}
\begin{equation}
\ell =\frac{\pi}{2} \sqrt{\frac{\kappa  T^{*}d }{J_c E}}
\left( 1 - \sqrt{\frac{2h_0 T^{*}}{nJ_c E}}\right)^{-1}\, .
\label{k_xy}
\end{equation}
Here,  $T^{*} \equiv - (\partial \ln J_c / \partial T)^{-1} $, 
$\kappa$ is the thermal conductivity, and $h_0$ is the coefficient of heat transfer between the film and the substrate.  
The  nonlinear current-voltage curve of the superconductor is described by the commonly used relation for the electrical field, $E \propto J^n$, where $n \gg 1$.
The threshold value for the applied perpendicular field, $H_{\text{th}}$, can then be found by combining Eq.~(\ref{k_xy}) with the Bean model expression for the flux penetration depth in a  thin strip\cite{brandt93, zeldov94} of half-width  $w$,  which gives,\cite{albrecht07}
\begin{equation}
H_{\rm th} =  \frac{J_c}{\pi} \;  {\rm arccosh} \left(\frac{1}{1-
\ell/w}
\right) .
\end{equation}

Plotted in Fig.~5 as a full curve is the generic relation between the  threshold  field and the critical sheet current based on the Eqs.(3) and (4). Included in the plot are also two line segments representing the values of the currents, $J_c^L$ and  $J_c^T$.
In zero in-plane field  the sample is isotropic, and the two lines overlap.
They are here drawn vertical, consistent with the Bean model approximation.

As pointed out above, when a  $H_{\parallel}$ is applied to the sample during the cooling, the resulting anisotropy is essentially due to  $J_c^T$ being reduced.
Thus, with increasing  $H_{\parallel}$ the dashed line  in Fig.~5  shifts to the left, while the full line representing $J_c^L$ remains essentially fixed.
Then, at some magnitude of $H_{\parallel}$,  the dashed line  enters the region where avalanches no longer will occur at a finite applied perpendicular field.
This explains the suppression of the thermomagnetic runaway as revealed in the image of Fig.~4(b). 

The suppression of thermomagnetic avalanches was observed also by placing a metallic layer nearby the superconducting film \cite{colauto_suppression_2010}.
Interestingly, this image shows that the incipient stage of the runaway has a flux distribution quite different from that created by a current pulse at 7~K.
The penetrated region has at 7~K the shape of a semicircle with radius close to  the width of the legs of the injector.\cite{colauto_anisotropic_2017}.
In Fig.~4(b) one sees a narrow protrusion with a width similar to that of the avalanche branches in panel (a).

\begin{figure}[t]
  \centering
  \includegraphics[width=8cm]{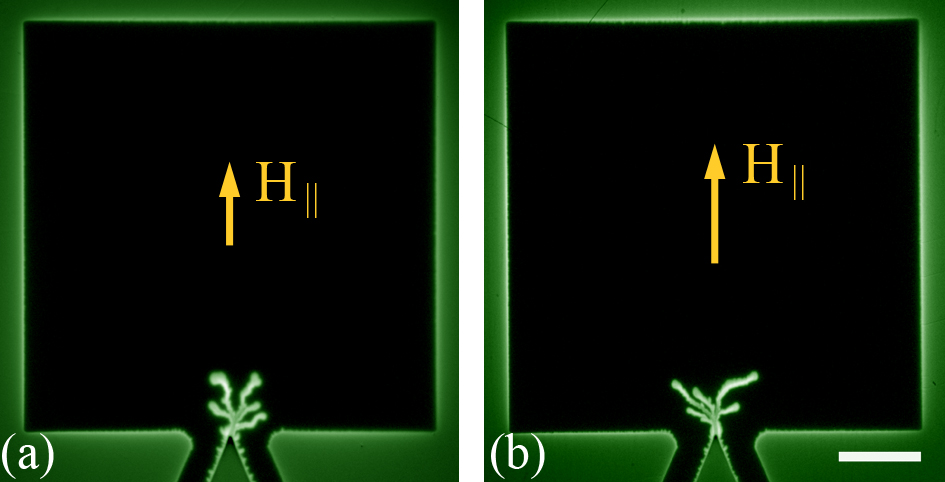}
  \label{ava-amplitude}
  \caption{Magneto-optical images of the Nb film with flux injector.
  In (a) and (b) the film was initially cooled to 2.5~K while applying in-plane fields (see arrows) of $H_{\parallel}$ = 0.7  and 1.0~kOe, respectively.
  The scale bar at the bottom right is 500~$\mu$m.
  }
\end{figure}

\subsection{Bending of dendrites}

Consider next current-induced  flux injections created for intermediate magnitudes of the frozen-in field $H_{\parallel}$. 
Shown in Fig.~6 (a) and (b) are magneto-optical images recorded at 2.5~K after the sample was cooled in the fields  0.7 and 1.0 kOe, respectively.
In both cases the current pulse resulted in a dendritic avalanche, and their overall size is seen to be similar to that in Fig.4~(a).
However, the images show a new feature, namely that  with increasing  $H_{\parallel}$ the avalanche branches tend to propagate by bending towards the direction perpendicular to the frozen-in field. Moreover, the degree of bending appears to increase with the strength of the field $H_{\parallel}$.

\begin{figure}[b]
  \centering
  \includegraphics[width=8cm]{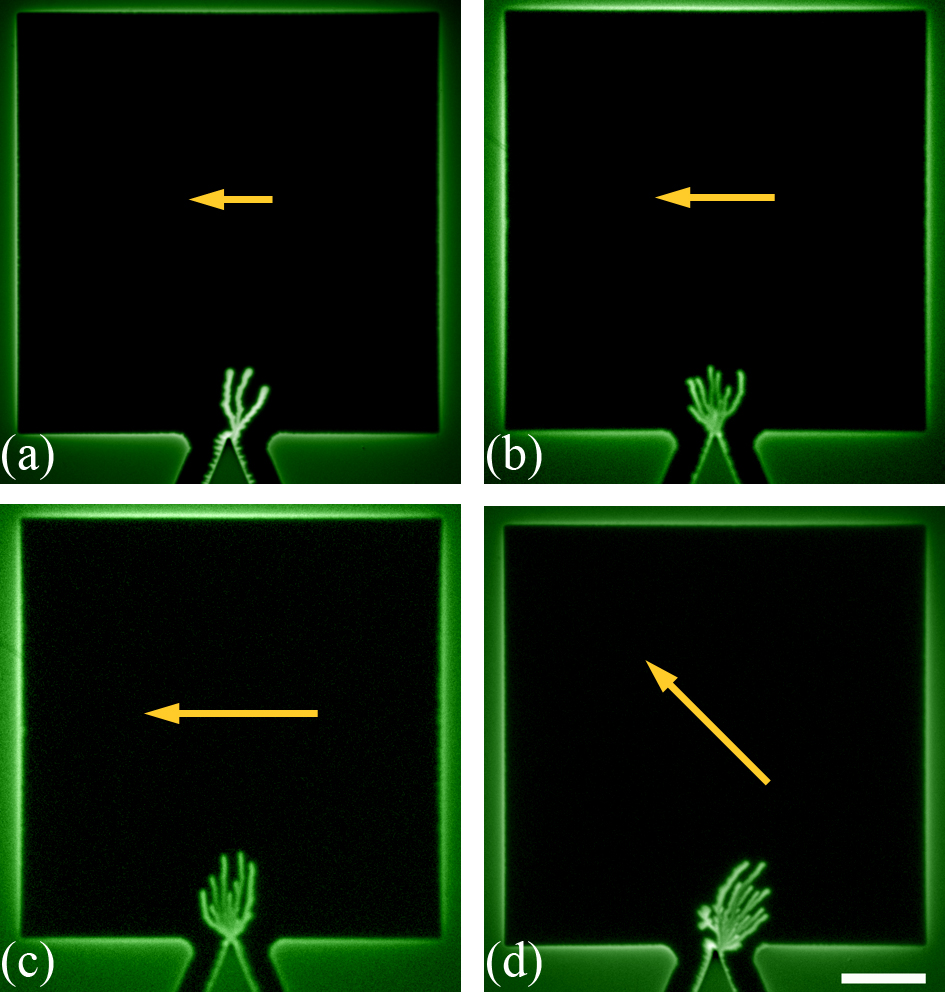}
  \label{ava-direction}
  \caption{Magneto-optical images of the Nb film with flux injector at 2.5~K.
  In panels from (a) to (c) the in-plane field was rotated by ${90}^{\circ}$, and the magnitudes are 0.7, 1.0, and 1.5~kOe, respectively.
  In the panel (d) the frozen-in in-plane field was rotated by ${45}^{\circ}$ and the magnitude is 1.5~kOe.
  The directions of the in-plane field are indicated by the arrows.
  The scale bar at the bottom right is 500~$\mu$m.
   }
\end{figure}

To investigate this effect further, in-plane fields  were frozen-in  using several other directions, and also applied with different magnitudes. Presented in Fig.~7 are typical images of avalanches triggered by the flux injector using again current pulses $i_{app} = 1$~A of duration 200~ms. 

In the panels  (a) - (c) of Fig.~7 the arrows show that the direction of  $H_{\parallel}$ was here rotated ${90}^{\circ}$ relative to $H_{\parallel}$ in Fig.~6. The length of the arrows indicates that the magnitude of $H_{\parallel}$ increases from panels (a) to (c).
By comparing these three images it is evident that the dendrites consistently tend to deflect towards the direction perpendicular to $H_{\parallel}$. 
 Moreover, one also observes that the deflection increases with the  magnitude of $H_{\parallel}$.
 
Interestingly, from the panels (a)-(c) in Fig.~7  and the zero-field image in Fig. 4(a), one sees a systematic trend that the length of each branch of these avalanches increases with the magnitude of  $H_{\parallel}$. 
Striking is also the image in panel (d) of Fig.~7, where the  $H_{\parallel}$ was rotated 45 degrees.
The avalanche branches clearly followed the field rotation, and display a pronounced elongation perpendicular to  $H_{\parallel}$. 

To shed some light on this  intriguing behavior we draw attention to the physical picture outlined in Ref.~\onlinecite{Vestgarden12}. 
There, it was identified three stages of a thermomagnetic avalanche in a superconducting film.
After an initial nucleation stage, there is a stage with rapid propagation, typically 100 km/s, \cite{Bolz2003, Vestgarden12} of a dendritic structure of magnetic flux and elevated temperature.
This is a stage governed by the nonlocal and non-linear electromagnetic response of the superconducting film.
Finally comes a retardation stage governed by cooling of the remaining active parts of the avalanche.
This is a process of lateral spreading of heat controlled by the thermal conductance in the film, and the heat flow into the substrate. 

We expect that this last stage is when the avalanche is most strongly influenced by the frozen-in in-plane vortices.
Assuming that the thermal conductance along the array of vortices is larger than the conductance perpendicular to them, cf. Refs.~{\onlinecite{Maki67,Maki69}, the cooling  becomes more efficient in the direction along $H_{\parallel}$. Since efficient cooling reduces the thermomagnetic feedback, propagation in the direction of small thermal conductivity will be favourable in fueling the runaway.
This qualitative argument largely explains the field-induced bending of dendrites reported in this work.
To provide a quantitative theoretical estimate of the field-induced anisotropy was not successful because an analytical theory of the 
developed stage of dendrite propagation is not available.

\section{Conclusions}

In conclusion, we have observed that cooling a Nb film in the presence of an in-plane magnetic field, $H_{\parallel}$, is a highly efficient way to create anisotropy, and that the anisotropy increases rapidly following a cubic dependence on $H_{\parallel} ~$.
By applying an active flux injector, it was found at low temperatures, where flux penetrates mainly in the form of thermomagnetic avalanches, that also this ultra-fast flux dynamics is strongly affected by the frozen-in $H_{\parallel}$.
With $H_{\parallel}$ reaching 1.5 kOe the avalanches stopped to occur, showing that there exists an externally controllable way to quench thermomagnetic runaways.
When applying smaller $H_{\parallel}$,  the branches of the avalanche consistently tend to bend towards the direction transverse to the frozen-in field.
The present results therefore demonstrate that by applying in-plane fields to a thin superconductor, one has a versatile external tool to control flux dynamics, 
and even prevent a thermomagnetically unstable region to develop into a large avalanching structure.

\section*{acknowledgments}

The samples were grown in Laborat\'{o}rio de Conforma\c{c}\~{a}o Nanom\'{e}trica (LCN-IF-UFRGS), and the lithography was made in Laborat\'{o}rio de Microfabrica\c{c}\~{a}o (LMF/LNNano/CNPEM).
The work was partially supported by the Sao Paulo Research Foundation (FAPESP) Grant No. 2013/16.097-3, the Brazilian National Council for Scientific and Technological Development (CNPq), the Brazilian program Science without Borders, as well as the CAPES-SIU-2013/10046 project \enquote{Complex fluids in confined environments}.

\bibliography{references}

\end{document}